\documentclass{aipproc}

\layoutstyle{8x11double}

\SetInternalRegister\hbadness{8000} % pseudo latin isn't breaking very well :-)

\begin{document}

\title 
      [In flight performance and first results of FREGATE]
      {In flight performance and first results of FREGATE}


\author{J-L. Atteia} 
{address={C.E.S.R., 9 Avenue du Colonel Roche, 31028 Toulouse
Cedex 4, FRANCE},   email={Jean-Luc.Atteia@ast.obs-mip.fr}}

\author{M. Boer} 
{address={C.E.S.R., 9 Avenue du Colonel Roche, 31028 Toulouse Cedex 4, FRANCE}}

\author{F. Cotin} 
{address={C.E.S.R., 9 Avenue du Colonel Roche, 31028 Toulouse Cedex 4, FRANCE}}

\author{J. Couteret} 
{address={C.E.S.R., 9 Avenue du Colonel Roche, 31028 Toulouse Cedex 4, FRANCE}}

\author{J-P. Dezalay}
{address={C.E.S.R., 9 Avenue du Colonel Roche, 31028 Toulouse Cedex 4, FRANCE}}

\author{M. Ehanno} 
{address={C.E.S.R., 9 Avenue du Colonel Roche, 31028 Toulouse Cedex 4, FRANCE}}

\author{J. Evrard} 
{address={C.E.S.R., 9 Avenue du Colonel Roche, 31028 Toulouse Cedex 4, FRANCE}}

\author{D. Lagrange} 
{address={C.E.S.R., 9 Avenue du Colonel Roche, 31028 Toulouse Cedex 4, FRANCE}}

\author{M. Niel} 
{address={C.E.S.R., 9 Avenue du Colonel Roche, 31028 Toulouse Cedex 4, FRANCE}}

\author{J-F. Olive} 
{address={C.E.S.R., 9 Avenue du Colonel Roche, 31028 Toulouse Cedex 4, FRANCE}}

\author{G. Rouaix}
{address={C.E.S.R., 9 Avenue du Colonel Roche, 31028 Toulouse Cedex 4, FRANCE}}

\author{P. Souleille} 
{address={C.E.S.R., 9 Avenue du Colonel Roche, 31028 Toulouse Cedex 4, FRANCE}}

\author{G. Vedrenne}  
{address={C.E.S.R., 9 Avenue du Colonel Roche, 31028 Toulouse Cedex 4, FRANCE}}

\author{K. Hurley}
{address={UC Berkeley Space Sciences Laboratory, Berkeley, CA 94720-7450}}

\author{G. Ricker}{
  address={M.I.T. Center for Space Research, 70 Vassar St., Cambridge, MA 02139}
}

\author{R. Vanderspek}{
address={M.I.T. Center for Space Research, 70 Vassar St., Cambridge, MA 02139}
}

\author{G. Crew}{
address={M.I.T. Center for Space Research, 70 Vassar St., Cambridge, MA 02139}
}

\author{J. Doty}{
address={M.I.T. Center for Space Research, 70 Vassar St., Cambridge, MA 02139}
}

\author{N. Butler}{
address={M.I.T. Center for Space Research, 70 Vassar St., Cambridge, MA 02139}
}

% \copyrightholder{Acoustical Scociety of America}
\copyrightyear  {2001}

\begin{abstract}
The gamma-ray detector of HETE-2, called FREGATE, has been designed to detect
gamma-ray bursts in the energy range [6-400] keV. Its main task is to
alert the other instruments of the occurrence of a gamma-ray burst (GRB) and to provide the
spectral coverage of the GRB prompt emission in hard X-rays and soft gamma-rays.
FREGATE was switched on on October 16, 2000, one week after the
successful launch of HETE-2, and has been continuously working since then.
We describe here the main characteristics of the instrument, its
in-flight performance and we briefly discuss the first GRB observations. 
\end{abstract}

\date{\today}

\maketitle
\section{Introduction}

The HETE-2 spacecraft (Ricker et al. 2002a, Doty et al. 2002) has been designed to
distribute gamma-ray burst localizations within several seconds of
the burst detection. 
The localization process and the distribution of rapid
alerts is a complex chain of events which starts when a GRB is detected
and identified as such. HETE-2 (hereafter HETE
for simplicity) carries three experiments: the Soft X-ray Camera (SXC), 
The Wide-Field X-ray Monitor (WXM) and the FREnch GAmma TElescope (FREGATE). 
The latter is a traditonal gamma-ray detector operating
in the energy range [6-400] keV, which was built
by the Centre d'Etude Spatiale des Rayonnements (CESR) in Toulouse, France.
FREGATE has three main goals:

\begin{itemize}
\item
Detecting count rate increases and qualifying them as gamma-ray
burst candidates.

 \item 
 Performing GRB spectroscopy over a broad energy range (complementing
and overlapping the WXM energy range).

\item
Monitoring the activity of galactic transients like the Soft Gamma
Repeaters. 
 
\end{itemize}

This paper describes FREGATE and explains how it is operated and
calibrated in-flight. It also contains a brief discussion of the types of
events detected during the first year of operation and a highlight of the
preliminary scientific results of the mission.

\section{Description of the instrument}

\begin{figure}
\resizebox{0.95\columnwidth}{!}
{\includegraphics{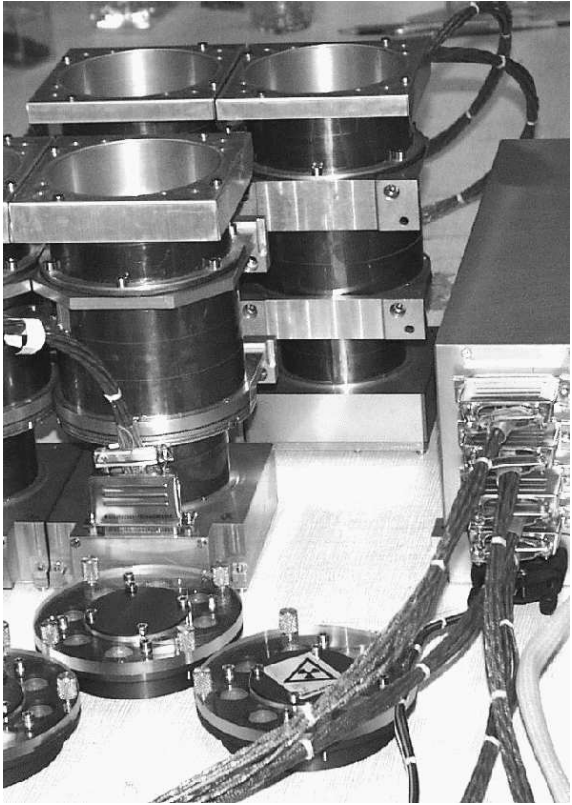}}   
%\spaceforfigure{75mm}{60mm}
\caption{FREGATE in the laboratory before its integration on the
spacecraft. The four detectors on the left are 20 cm high.} 
\label{labo981209}
\end{figure}

The FREGATE hardware consists of four identical detectors and one electronics
box (see Figure \ref{labo981209}). The instrument weights 14 kg and consumes 9
watts of electrical power. The four detectors are co-aligned on the spacecraft
in order to share the same field of view. Contrary to the other HETE
instruments (the WXM and the SXC), FREGATE has no localization
capability, except the ability to recognize whether a transient occurred within
or outside its field of view. The main characteristics of FREGATE are
described in table \ref{perfo}.

\begin{table}
\begin{tabular}{lrr}
\hline
Energy range & 6 - 400 keV \\
Effective area (4 detectors, on axis) & 160 cm$^2$ \\
Field of view (FWZM) & 70$^\circ$ \\
Sensitivity (50 - 300 keV) & 10$^{-7}$ erg cm$^{-2}$ \\
Dead time &  10 $\mu sec$ \\
Time resolution & 6.4 $\mu sec$ \\
Maximum acceptable photon flux & 10$^3$ ph cm$^{-2}$ sec$^{-1}$\\
Spectral resolution at 662 keV & $\sim$ 8 \%  \\
Spectral resolution at 122 keV & $\sim$ 12 \%  \\
Spectral resolution at 6 keV & $\sim$ 42 \%  \\
\hline
\end{tabular}
\caption{FREGATE performances}
\label{perfo}
\end{table}

\begin{figure}
\resizebox{0.95\columnwidth}{!}
{\includegraphics{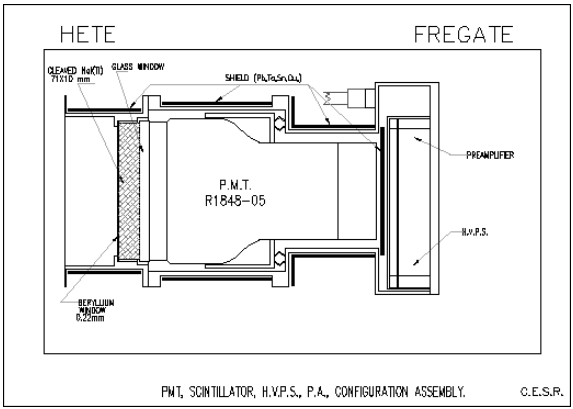}}
\caption{Schematic drawing of one FREGATE detector. The crystal has a
diameter of 71mm and a thickness of 10mm. The length of the detector is 20cm.} 
\label{hete102} \end{figure}

\subsubsection{Detectors} 

\begin{figure}
\resizebox{0.95\columnwidth}{!}
{\includegraphics{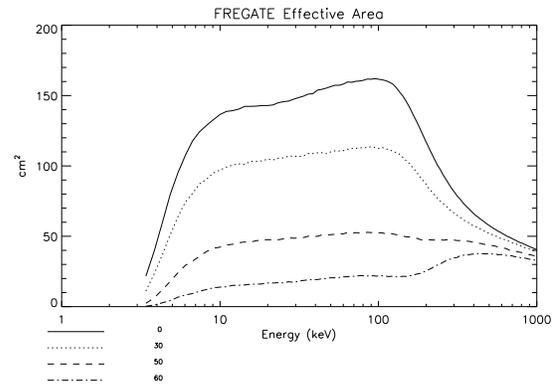}}   
\caption{FREGATE effective area for on-axis and off-axis sources.}
\label{effective_area}
\end{figure}

Each FREGATE detector consists of a cleaved NaI crystal (a cylinder 10mm thick
and 71mm in diameter) read by a photomultiplier (Hamamatsu 1848, figure
\ref{hete102}). In order to extend the spectral coverage of FREGATE at 
low energies, we chose cleaved crystals which have no dead layer. 
For the same reason the crystals are encapsulated in a beryllium housing
which reduces the absorption of low energy gamma-rays. The thickness of
the housing on the front side of the detectors is 0.22 mm and
the overall transmission of the entrance window is larger than
65\% at 6 keV and reaches 85\% at 10 keV (see figure \ref{effective_area}).
At high energies the electronics limits the energy range to photons with
energies below 400 keV. 
The geometric area of the sum of the four detectors for a source on-axis is
nearly 160 cm$^2$. 
In addition there are two on-board  radioactive sources of Barium 133 which
illuminate the detectors from the outside with photons at 81 and 356 keV, allowing to monitor
the gain of the detectors from the ground (see subsection "In-flight calibration"
for more details) .

\subsubsection{Shield and Collimator}

As shown in figure \ref{hete102}, the body of the detectors is surrounded by a
graded shield made of lead, tantalum, tin, copper, and aluminium (0.8 mm of lead, 0.3 mm of
tantalum, 0.7 mm of tin, 0.3 mm of copper, and 0.8 mm of aluminum). The aim of
the shield is to prevent the photons from outside the field of view from reaching the
crystal. The transparency of the shield is 5.5\% at 150 keV, 55\% at 300 keV, and
75\% at 500 keV.

\begin{figure}
\resizebox{0.95\columnwidth}{!}
{\includegraphics{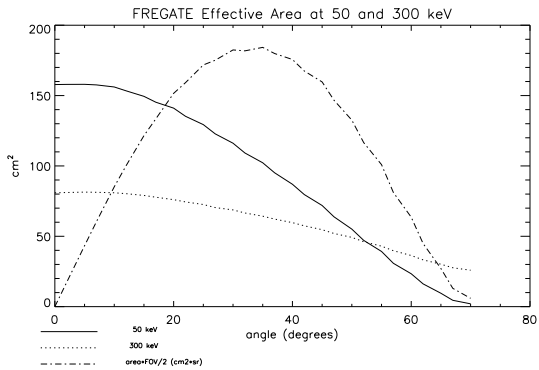}}   
\caption{FREGATE response to off-axis sources. 
Solid line: angular response at 50 keV where the collimator is opaque. 
dotted line: angular response at 300 keV where the collimator is mostly transparent. 
Dash-dotted line: expected number of GRBs as a function of their angle of arrival.}
\label{area_ang}
\end{figure}

A peculiarity of the FREGATE detectors is that the shield goes beyond the
front face of the crystal, reducing the field of view (FOV) of the instrument
and acting as a collimator. The angular response of FREGATE with its collimator
is shown in figure \ref{area_ang}. While this collimator decreases the
number of GRBs that FREGATE can detect, it plays an essential role for a
mission like HETE where a synergy must be found between three sets of
instruments with different properties and constraints. In the context of
FREGATE the collimator provides the following advantages (in order of
increasing importance):

\begin{itemize}
\item
It increases the fraction of FREGATE GRBs which are within the FOV of the WXM. 

\item 
It decreases by a factor of two the count rate from the diffuse
X-ray background.

\item
It restricts to 4 months per year the transit time of galactic sources within
the field of view of FREGATE (instead of nearly 6 months).

\end{itemize}

While GRBs which are outside the field of view of the WXM cannot be localized,
it was thought that the FOV of FREGATE should nevertheless be wider than the
FOV of the WXM. The main reason for this choice was that a wider field of view 
of FREGATE ensured that all the GRBs detected by the WXM would illuminate at
least 60 cm$^2$ of FREGATE detectors (as was the case for GRB010921, Ricker et al 2002b). 
Moreover FREGATE-only GRBs are useful for broadband spectroscopic studies, 
they increase the statistics for rare events (like X-Ray Flashes or short
GRBs) and they can be localized by the IPN.

\subsubsection{Readout electronics} 
Each detector has its own analog and digital electronics. 
The analog electronics contain a discriminator circuit with four adjustable
channels and a 14-bit PHA whose output is regrouped into 512 evenly-spaced
energy channels (each approximately 0.8 keV wide). The (dead) time needed to
encode the energy of each photon is 14 $\mu$s for the PHA and 9 $\mu$s for
the discriminator.
The digital electronics process the individual pulses, to produce
the following output : 

\begin{itemize}
\item
Time histories in 4 energy channels (every 20ms). 

\item 
128-channel energy spectra spanning the range 0-400 keV (every 80 ms).

\item
A circular buffer containing the most recent 65536 photons tagged in time
(resolution 6.4 $\mu$s) and in energy (resolution 0.8 keV)..

\end{itemize}

\subsubsection{On-board software}

The main tasks of the on-board software include the configuration of the
instrument, the acquisition of data from the electronics, their packaging
for the telemetry, and the search for excesses in the count rate.

{\it Configuration of the instrument.}The output of FREGATE depends on a number
of adjustable parameters such as the settings of the high voltages, the limits of
the energy channels, the trigger criteria, or the on-board data compression. 
These parameters are described in a configuration file which can be uploaded
when the spacecraft is in contact with one of the three primary ground stations
(PGS, see Crew et al. 2002). When a new configuration file is
uploaded, the on-board software modifies the configuration of FREGATE
accordingly.

{\it Data packaging.} Every ~20 ms the FREGATE Digital Signal Processor (DSP)
reads the data from the electronics and prepares the data for the telemetry.
The data products generated by FREGATE are described in the next Section.

{\it Search for excesses.} The on-board software also scans the data
in real time to search for sudden increases in the count rate recorded by the
instrument. This work is described in details in Section "FREGATE
triggers" below. 

\section{Data Types}

\begin{figure}
\resizebox{0.95\columnwidth}{!}
{\includegraphics{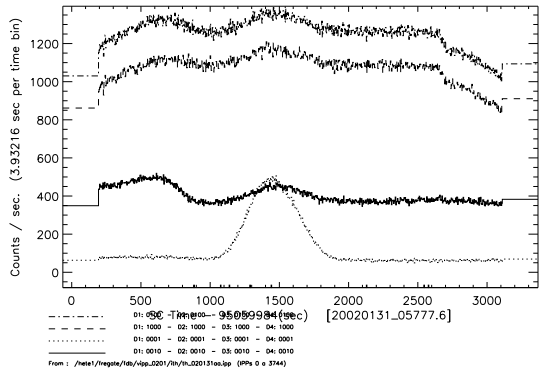}}   
\caption{Continuous data generated by FREGATE during nighttime. From top to
bottom, the light curves in the 4 energy bands B, A, C, and D (see text) for
the sum of the 4 detectors. The data have been regrouped in 4 second bins for
clarity, the actual resolution of the data is 0.16 seconds. The large peak
in energy band D is due to protons trapped in the SAA.}   
\label{continuouslc}  
\end{figure}

\begin{figure}
\resizebox{0.95\columnwidth}{!}
{\includegraphics{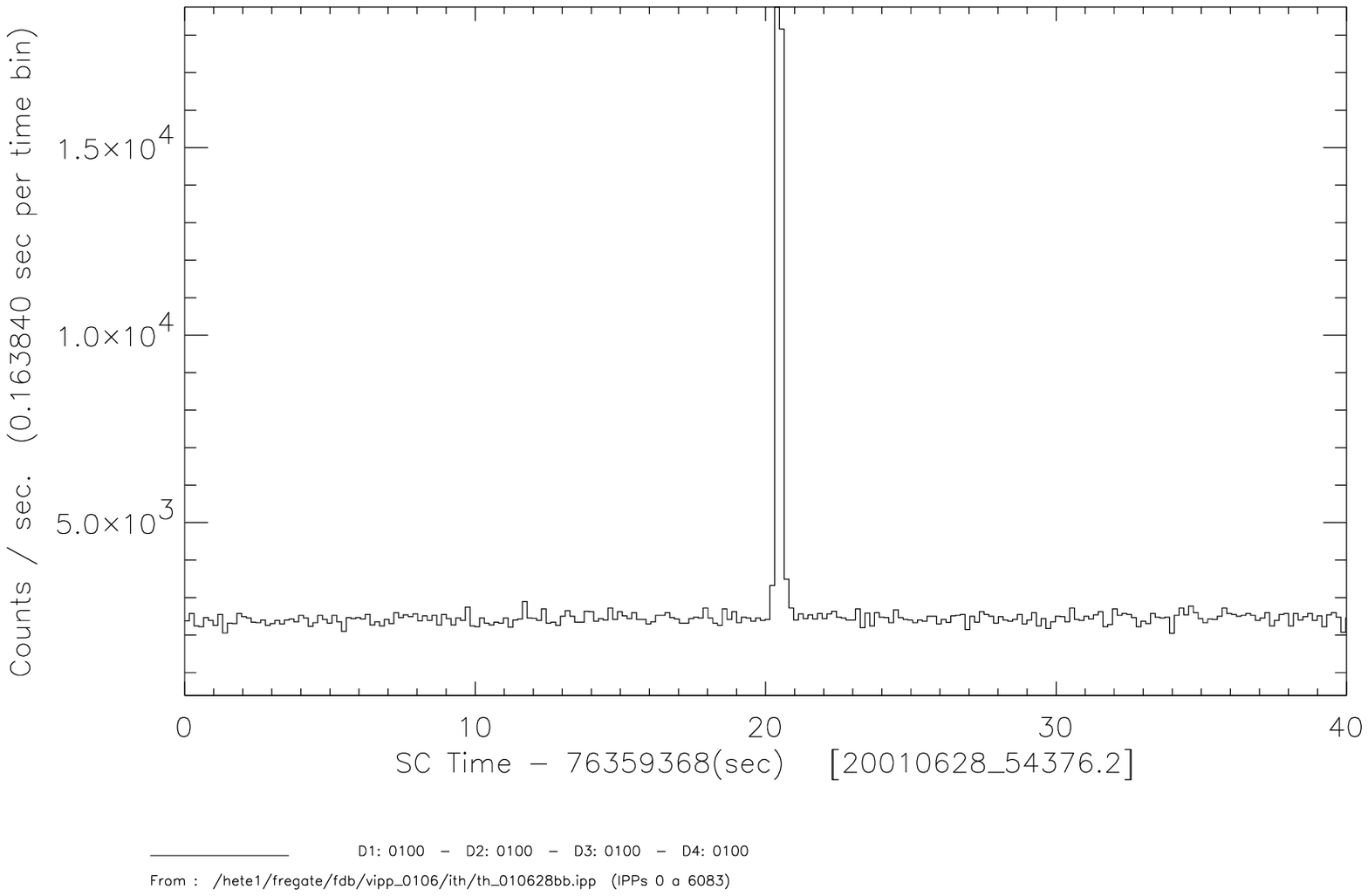}}
\resizebox{0.95\columnwidth}{!}
{\includegraphics{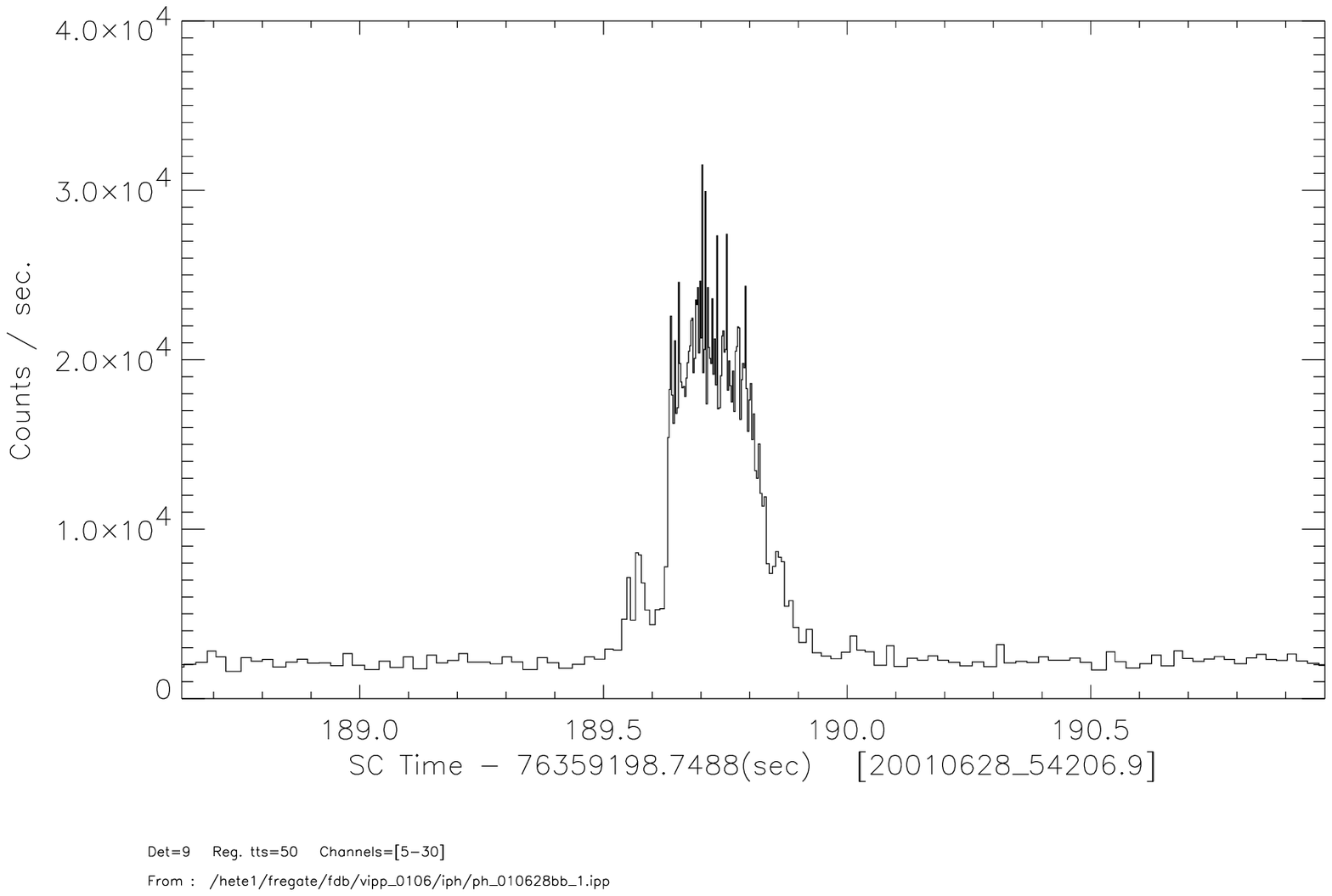}}   
\caption{A burst from SGR 1900+14 as seen in the continuous data (left) and in the
burst data (right).} 
\label{burstdata}
\end{figure}

FREGATE generates 4 types of data:
Housekeeping (HK), light curves and spectra generated by the FREGATE DSP, light
curves generated by one of the on-board tranputers (the so-called X-$\gamma$ 
transputer), and burst data.
Housekeeping data are produced continuously; light curves and spectra are produced
when the high voltages are on; and burst data are produced only after a
trigger.

The HK data and the light curves produced by the on-board transputer will not
be discussed here, however, the continuous data and the burst data generated by the
FREGATE DSP are explained in detail below.

\subsubsection{Light curves and spectra generated by the DSP}
During nighttime, when the high voltages are on, FREGATE produces continuous
light curves and spectra. 

The light curves represent the count rates measured by the 4 detectors
in 4 broad energy channels with time resolutions of 0.16 and 0.32 s. 
The limits of the energy channels are usually set to 

\begin{itemize}
\item
6 - 40 keV for channel A. 

\item 
6 - 80 keV for channel B. 

\item
32 - 400 keV for channel C. 

\item
> 400 keV for channel D. 

\end{itemize}

An example of these data is shown in figure \ref{continuouslc}.

Simultaneously FREGATE generates four 128-channel energy spectra covering the
energy range 0-400 keV every 5 or 10 seconds (but the electronics threshold
and the absorption of the beryllium window reduce the effective energy range
to 6-400 keV).

\subsubsection{Burst data}
When a trigger occurs, burst data are generated in addition to the continuous
data. The burst data consist of 256k photons (64k per detector) tagged in time
(with a resolution of 6.4 $\mu$s) and in energy (256 energy channels spanning
the range 0-400 keV). These burst data allow detailed studies of the
spectro-temporal evolution of bright GRBs. An example of the gain provided by
the burst data is shown in figure \ref{burstdata}.

\section{FREGATE operation}

FREGATE operations are driven by alternating nighttime and
daytime periods. Because HETE instruments always point in the antisolar
direction they have the earth in their field of view during about 45 min per
orbit (the duration of one orbit is 90 min); this is the daytime period.

During daytime the High Voltages of the detectors are switched off, partly to
reduce the amount of data produced by the spacecraft and partly to avoid
the triggers due to solar X-ray flares reflected on the atmosphere of the Earth
(see figure \ref{solarflare}). 

During nighttime the high voltages are switched on. The detectors
continuously record the gamma-ray flux in four energy bands and these data are
processed by the DSP to search for excesses due to GRBs (see next section). 

\subsubsection{In-flight calibration}

Since FREGATE records only the time and the energy of the photons, we 
need to calibrate only the FREGATE timing and energy scales. 

The FREGATE time is directly derived from the HETE time and the
calibration of FREGATE time is, in reality, the issue of the spacecraft
time calibration which is not discussed here.

\begin{figure}
\resizebox{0.95\columnwidth}{!}
{\includegraphics{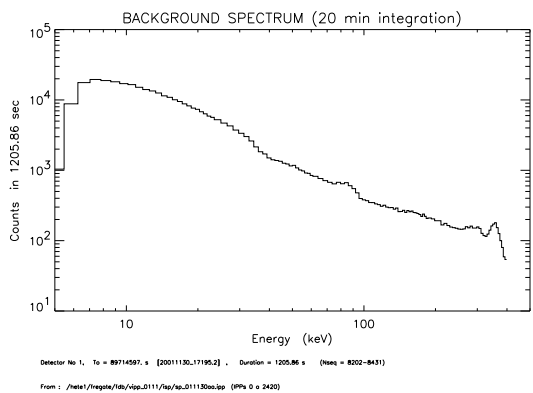}}   
\caption{Background spectrum showing the two calibration lines at 81 keV and
356 keV.} 
\label{bdf_wh}
\end{figure}

As mentioned above, the energy calibration of the detectors is made possible by two
on-board radioactive sources of Barium 133 which illuminate the detectors
from the outside ($^{133}$Ba emits gamma-ray lines at 81 and
356 keV, with a half-life of 10.5 years) . 
A typical background spectrum, accumulated for 1200
seconds, is shown in figure \ref{bdf_wh}. These radioactive sources allow the
monitoring of the gain of the four detectors of FREGATE with a time resolution of
a few minutes (the gain is manifested in the response of the detectors to photons with a
given energy). The gain fluctuates on two timescales: along one orbit (90 min)
and on a longer period of several weeks. 

\begin{figure}
\resizebox{0.95\columnwidth}{!}
{\includegraphics{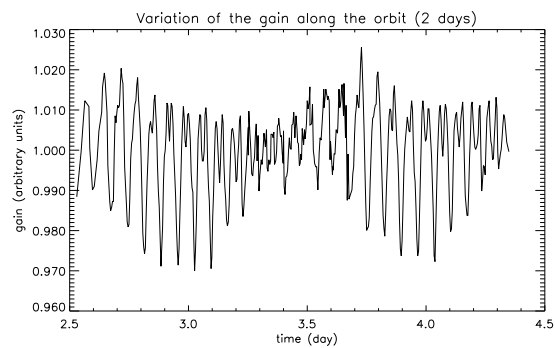}}   
\caption{Gain fluctuations during 26 consecutive orbit.
The gain fluctuations are due to the changing orientation of the magnetic field
along the orbit.}
\label{var_gain}
\end{figure}

On the short term,
the gain changes with the orientation of the magnetic field along the orbit.
This effect is due to the incomplete collection of the photoelectrons on the
first anode of the PMT (which has no magnetic shield). A typical example of
such variations is shown in figure \ref{var_gain}.

The gain also exhibits a tendency to decrease on the long term. this trend is
monitored on the ground and compensated by regular increases of the high
voltages (by a few percent) every 2 months.

\subsubsection{Energy response}

The in-flight energy response of a gamma-ray detector is the
combination of its physical response, its intrinsic
non-linearities, and the gain variations.
The physical response of FREGATE detectors has been evaluated with
detailed Monte Carlo simulations. The output of the simulation program has
been checked against ground calibrations made with 9 radioactive sources
having energies in the range 8 keV ($^{65}$Zn) to 1332 keV ($^{60}$Co). 
The same sources provide a measure of the intrinsic non-linearities of the
detectors. Finally the gain fluctuations are measured on-board as explained
above.

\begin{figure}
\resizebox{0.95\columnwidth}{!}
{\includegraphics{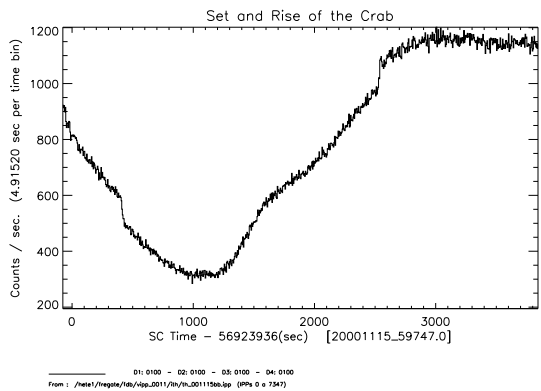}}   
\caption{Crab occultation: the two steps in the light curve are due to the
transit of the Earth in front of the Crab nebula. The light curve, which has a
temporal resolution of 5s, shows the total count rate measured by the four
FREGATE detectors in the energy range 6-80 keV. The Crab is approximately
30$^{\circ}$ off-axis.} 
\label{thecrab}
\end{figure}

The quality of the spectral response of FREGATE has been evaluated via the
deconvolution of the hard X-ray emission from the Crab nebula. The spectrum of
the Crab nebula has been constructed from the amplitude of the Crab occultation
steps observed at various energies, see figure \ref{thecrab}. 
The deconvolved spectrum is fully compatible with the well known spectrum of
the Crab nebula for energies between 10 keV and 200 keV and for angles between
0$^{\circ}$ and 50$^{\circ}$. This procedure and the results obtained are
described in details by Olive et al. (2002a).

\subsubsection{Ground operations}
The operational tasks on the ground are very simple. 
In normal operation they include the following actions :

\begin{itemize}
\item
Upload the HV ON/OFF sequences.

\item 
Check the health of the detectors.

\item
Adjust the gains and update the configuration.

\item
Search for events which didn't fire the on-board trigger.

\end{itemize}

When an astronomical event is detected it is also necessary to update the
FREGATE catalog(s) and to construct the response matrices if
it is within the field of view.

\section{FREGATE Triggers}

\subsubsection{GRB detection}
The data recorded by FREGATE are searched for GRBs and other astronomical
transients both on-board and on the ground. Two real-time programs run on-board:
the DSP trigger and the transputer trigger. The former is
described below. The transputer trigger has been designed to add more
flexibility and to search for excesses in the combined data from FREGATE and
the WXM, it is described by Tavenner et al. 2002.

The on-board GRB detection is completed by two programs which
automatically process the data when they arrive at the ground. These
programs are more efficient than the on-board processing for the detection of long 
or soft events, they are described in Butler et al. 2002 and in Graziani et al. 2002.

\paragraph {DSP trigger}
The DSP triggers when the count rate measured over a time interval
$\Delta t$ exceeds the average count rate, measured over the last $T$ seconds,
by more than $k$ standard deviations. 
Four timescales ($\Delta t$) are used for the trigger detection : 20 ms, 160
ms, 1.3 s, and 5.2 s.
The duration of the background integration is an adjustable parameter, usually
set to 30 s. 
The trigger thresholds $k$ are adjustable and currently set to values between
4.5 and 6.
The trigger detection algorithm works in parallel in the energy channels B
(6-80 keV) and C (30-400 keV). In order to decrease the rate of false triggers,
due to electronic noise or particles, we discard the triggers which are detected 
by only one of the four detectors. 
When the DSP triggers it sends an alert message to the HETE trigger monitor
(Crew et al. 2002, Vanderspek et al. 2002),
which alerts the other instruments of HETE and the VHF transmitter.
Simultaneously FREGATE starts to record burst data (see "DATA TYPES").
Additionally,
FREGATE will go into burst mode when it receives a message from the on-board
trigger monitor (e.g. following a WXM trigger).
Because the high-energy sky is essentially variable, many types of events
can trigger FREGATE. We summarize below the origin of the triggers
detected during the first year of the mission.

\begin{figure}
\resizebox{0.95\columnwidth}{!}
{\includegraphics{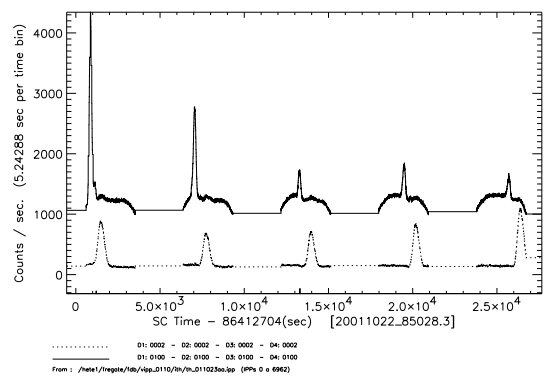}}   
\caption{Electrons and protons trapped in the radiation belts. The lower
curve shows the signal due to the protons in the South Atlantic Anomaly.
The upper curve shows large peaks due to trapped electrons, they are detected
a few minutes before the SAA.} 
\label{presaa}
\end{figure}

\subsubsection{Non-astrophysical events}

\paragraph{Pre-SAA electrons} 
In addition to the high fluxes of protons detected in the South Atlantic Anomaly (SAA),
energetic electrons are sometimes trapped in electron radiation belts crossed by HETE
a few minutes before the SAA. 
These populations of energetic electrons are highly variable and they reach 
a maximum in the days following large coronal mass ejections (CMEs) from
the Sun. When HETE goes through these rediation belts (at longitudes between 80 and
100 degrees) FREGATE measures high count rates due the interaction of the
electrons with the spacecraft (producing X-rays) and with the detectors (see figure \ref{presaa}) . 
These high count rates will sometimes trigger FREGATE.

\begin{figure}
\resizebox{0.95\columnwidth}{!}
{\includegraphics{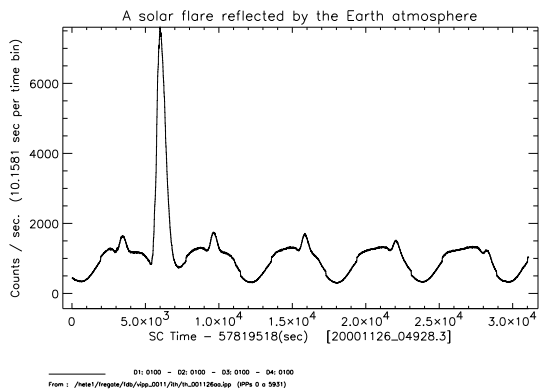}}   
\caption{A solar flare reflected by the atmosphere of the Earth on November 26 2000.}
\label{solarflare}
\end{figure}

\paragraph{Solar flares reflected on the Earth's atmosphere} FREGATE high voltages
are usually switched off during daytime. However, there are occasions when we want
FREGATE on with the Earth in the field of view. In this case the
low energy threshold of FREGATE make it very sensitive to solar flares
reflected by the atmosphere of the Earth. An exemple of such a flare (class
M2.2) is shown in figure \ref{solarflare}.

\paragraph{Noise triggers} The noise triggers are due to the statistical
fluctuations of the background count-rate measured by the detectors.
FREGATE triggers only when it detects two simultaneous excesses in the sum of
detectors 1+2 AND in the sum of detectors 3+4, this strategy reduces
the number of noise triggers to less than 1 per month.

\subsubsection{GRBs and other astrophysical transients}

\begin{figure}
\resizebox{0.95\columnwidth}{!}
{\includegraphics{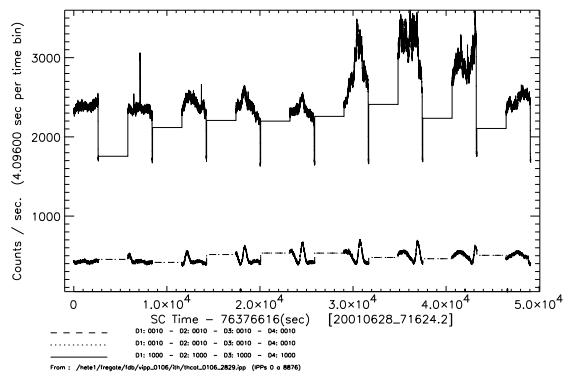}}   
\caption{Count rates measured by FREGATE in two energy channels (top: 6-40
keV; bottom 30-400 keV) during 9 consecutive orbits with SCO X-1 in the the field of view
(45 degrees off-axis). The flaring activity of SCO X-1 is clearly visible
during orbits 6 to 9 in the low energy channel. Note also that SCO X-1 has no effect
on the high energy channel (above 30 keV), which can be used to detect GRBs.
The three short spikes in orbit 2 and 3 are X-ray bursts.}   
\label{sco}
\end{figure}

\paragraph{\bf SCO X-1} SCO X-1 is the brightest hard X-ray source in the sky. It
exhibits rapid flaring (see figure \ref{sco}) and generates many triggers
when the trigger is enabled in energy channel B (6-80 keV). From the beginning of April
to the end of July, when SCO X-1 is within the field of view of FREGATE, the
trigger is disabled in channel B, completely suppressing SCO triggers.

\begin{figure}
\resizebox{0.95\columnwidth}{!}
{\includegraphics{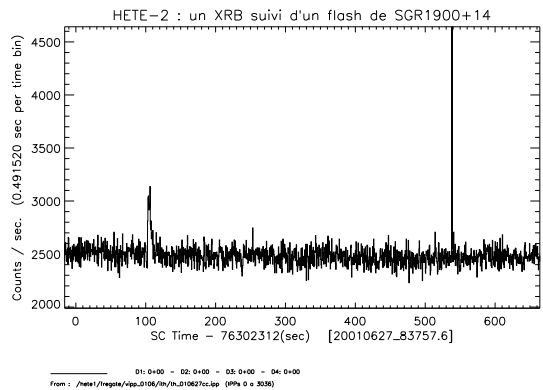}}
\caption{During summer time, when it has the galactic bulge within its
field of view, FREGATE detects many galactic transients. This light
curve (6-80 keV) shows an X-Ray Burst followed by a burst from SGR1900+14 about
seven minutes later.} 
\label{xrb_et_sgr} 
\end{figure}

\paragraph{X-ray bursts} X-ray bursts (XRBs) are due to thermonuclear
explosions at the surface of accreting neutron stars in binary systems.
Several dozen X-ray bursters are present in the galactic bulge. When FREGATE
has the galactic bulge within its field of view it detects a few XRBs per
day. These events do not trigger FREGATE, whose low energy trigger is disabled
during summer time (when SCO X-1 is in the field of view, see above), but they
are identified a posteriori by the ground processing. A sure way to identify
XRBs is to associate their arrival direction with a known X-ray source. The
WXM is well suited to do this job, but its field of view is only half the
field of view of FREGATE, implying that the identification of the XRBs
detected by FREGATE at large off-axis angles must rely solely on their
spectro-temporal properties. An example of an XRB detected by FREGATE can be
found in figure \ref{xrb_et_sgr}. A list of X-Ray Bursts detected and
localized by the WXM during the summer 2001 is given in Sakamoto et al. 2002.

\paragraph{Soft Gamma-ray Repeaters}
During the summer 2001, both SGR1900+14 and SGR1806-20 were active. From the
beginning of June to the end of August FREGATE detected about 30 short bursts
which can be attributed to these Soft Gamma Repeaters (figures
\ref{burstdata} and \ref{xrb_et_sgr}). Six of these bursts were localized by
the WXM (see Kawai et al. 2002), two were emitted by SGR1806-20 and four by
SGR1900+14. On July 2 2001, FREGATE detected a high
fluence burst from SGR1900+14 (GCN1078), a detailed analysis of this event is
given in Olive et al. (2002b).

\begin{figure}
\resizebox{0.95\columnwidth}{!}
{\includegraphics{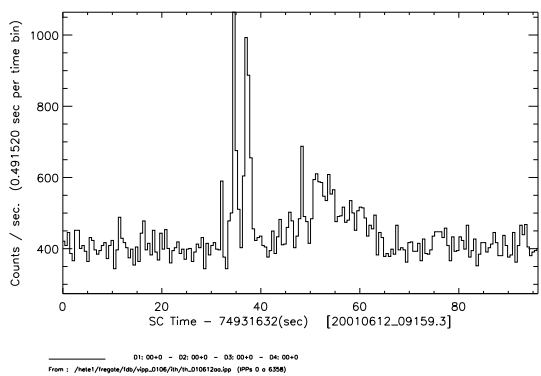}}
\caption{GRB010612 observed by FREGATE (30 to 400 keV), the light curve shows
the total of the four detectors, the GRB is 13 degrees off-axis.} 
\label{grb010612}
\end{figure}

\begin{figure}
\resizebox{0.95\columnwidth}{!}
{\includegraphics{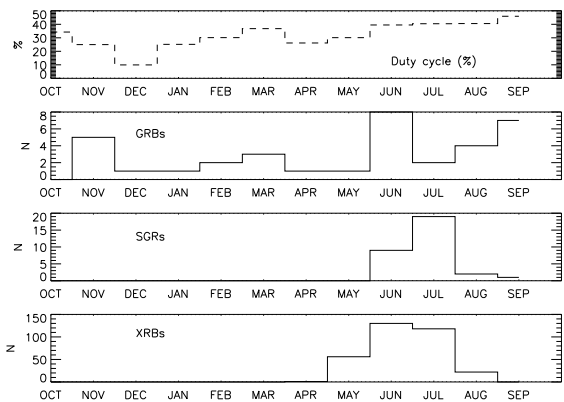}}   
\caption{Monthly detection number of various types of high energy transients from
October 2000 to September 2001. From May to August, the Galactic bulge was in 
the field of view of FREGATE. Note the ordinate of the XRB plot.} 
\label{nombres}
\end{figure}

\paragraph{Gamma-Ray Bursts}
Gamma-ray bursts constitute the main scientific target of HETE. Between October
2000 and September 2001, FREGATE has detected 32 confirmed GRBs and a
few unconfirmed events (see Dezalay et al. 2002 for a list). 
We estimated the sensitivity of FREGATE to be $10^{-7}$ erg cm$^{-2}$ in the energy
range [50-300] keV (Dezalay et al., 2002).
Figure \ref{nombres} shows the number of confirmed GRBs
detected by FREGATE since it was turned on in October 2000. The higher
number of GRBs per month after May 2001 is due to an increased observational
efficiency, which should result in the detection of a larger number of GRBs in
2002. 
Some interesting results have already been obtained with the GRBs detected during the first
year of FREGATE operation, the reader will find some of them in these proceedings and a short list
is given in the Conclusion below.

\section{Conclusion}
The first year of FREGATE operation shows that it fulfills the goals for which
it has been designed : alerting HETE when a GRB occurs, performing the
broadband spectroscopy of GRBs and other astrophysical transients (SGRs and
XRBs), and providing a census of galactic and extragalactic high energy
transients. Regarding this last issue, we note that, as a consequence of the
anti-solar pointing strategy of HETE, FREGATE observes the same portion of the
sky during more than $3 \times 10^6$ seconds per year. In addition FREGATE has been
succesfully integrated into the Interplanetary Network of gamma-ray burst
detectors (Hurley et al. 2002).

This volume contains some significant scientific results obtained by
HETE and FREGATE. 
The followup of GRB010921 detected by FREGATE and localized
with the WXM has led to the identification of the first HETE afterglow at a
redshift z=0.45 (Ricker et al. 2002b, Price et al., 2002). 
The spectro-temporal evolution of a bright burst from SGR1900+14 in analysed
in detail by Olive et al. (2002b). 
Barraud et al. (2002) discuss the existence of very soft GRBs (probably
similar to the X-Ray Flashes discussed by Heise, 2002) which have less than
10\% of their fluence above 30 keV.

\begin{theacknowledgments}
J-L Atteia acknowledges the support of the FREGATE team in CESR
(F. Cotin, J. Couteret, M. Ehanno, J. Evrard, D. Lagrange, G. Rouaix and
P. Souleille) who built a successful instrument and never lost enthusiasm after
the loss of HETE-1. A special tribute is due to M. Niel, who designed and
tested the prototype of the detectors. FREGATE is supported in France by the
CNES under contract CNES 793-01-8479.  None of the results presented here would
have been obtained without the dedication of the HETE OPS team at MIT who
manages FREGATE operations.  

\end{theacknowledgments}

% choose bibtex style depending on layout style and options used in
% sample:


\begin{thebibliography}{widest-label}

\bibitem{barraudwh} Barraud C., et al., these proceedings

\bibitem{butlerwh} Butler N., et al., these proceedings

\bibitem{crewwh} Crew G., et al., 2002, these proceedings

\bibitem{dezalaywh} Dezalay J-P., et al., 2002, these proceedings

\bibitem{dotywh} Doty J., et al., 2002, these proceedings

\bibitem{grazianiwh} Graziani C., et al., 2002, these proceedings

\bibitem{heisewh} Heise J., et al., 2002, these proceedings

\bibitem{hurleywh} Hurley K., et al., 2002, these proceedings

\bibitem{kawaiwh} Kawai N., et al., 2002, these proceedings

\bibitem{olivewh1} Olive J-F, et al., et al., 2002a, these proceedings 

\bibitem{olivewh2} Olive J-F., et al., et al., 2002b, these proceedings 

\bibitem{price02} Price. P., et al., 2002, submitted to ApJ

\bibitem{rickerwh} Ricker G., et al., 2002a, these proceedings

\bibitem{ricker02} Ricker G., et al., 2002b, submitted to ApJ

\bibitem{sakamotowh} Sakamoto T., et al., 2002, these proceedings

\bibitem{tavennerwh} Tavenner T., et al., 2002, these proceedings

\bibitem{vanderspekwh} Vanderspek R., et al., 2002, these proceedings

\end{thebibliography}
\end{document}